\documentclass{appolb}
\usepackage{epsfig}
\usepackage{amsmath}
\usepackage{slashed}

\begin{document}
\title{Probing the nature of phases across the phase transition 
at finite isospin chemical potential
\thanks{Presented by Rajiv V. Gavai at Critical Point and Onset of Deconfinement
CPOD2016, Wroc\l{}aw, Poland}
}
\author{Gunnar S. Bali 
\address{Institut f\"ur Theoretische Physik, Universit\"at Regensburg, 
Universit\"atsstra\ss{}e 31, 93053 Regensburg, Germany}
\and
G. Endr{\H o}di
\address{Institute for Theoretical Physics, Goethe University,
Max-von-Laue-Strasse 1, 60438 Frankfurt am Main, Germany and Institut f\"ur
Theoretische Physik, Universit\"at Regensburg, Universit\"atsstra\ss{}e 31,
93053 Regensburg, Germany }
\and
Rajiv V. Gavai \& N. Mathur
\address{Department of Theoretical Physics, Tata Institute of Fundamental 
Research, Homi Bhabha Road, Mumbai 400005, India
}}
\maketitle
\begin{abstract}

We compare the low eigenvalue spectra of the Overlap Dirac operator on two sets
of configurations at $\mu_I/\mu_I^c$ = 0.5 and 1.5 generated with dynamical
staggered fermions at these isospin chemical potential on $24^3 \times 6$
lattices.  We find very small changes in the number of zero modes and low lying
modes which is in stark contrast with those across the corresponding finite
temperature phases where one sees a drop across the phase transition. Possible
consequences are discussed.

\end{abstract}


\section{Introduction}

The baryon density-temperature ($\mu_B$-$T$) phase diagram of quantum
chromodynamics (QCD) has received a lot of attention for the past few decades,
starting from skeleton diagrams on the basis of simple hadronic models, which
explain the hadron spectrum reasonably well, to the increasingly quantitative
attempts to pin it down {\it ab initio} from QCD itself using the
non-perturbative lattice approach. As is well-known, one has to face the famous
fermion sign(phase) problem at nonzero baryon density or equivalently nonzero
baryon chemical potential, $\mu_B$, adding an extra layer of uncertainty to the
results obtained. In addition to baryon number, the up and down quarks also
carry isospin.  Defining $\mu_I$ as the chemical potential for $I_z$, and
$\mu_u$, $\mu_d$ for the up and down quarks, one has $\mu_B = 3 (\mu_u +
\mu_d)/2$ and $\mu_I = (\mu_u - \mu_d)/2$ or alternatively, $\mu_u=\mu_B/3 +
\mu_I$ and $\mu_d =\mu_B/3 -\mu_I$. The fermion determinant is real \cite{ss,ks}
for $\mu_I \ne 0$ and $\mu_B =0$, and one thus has no sign problem in that case.
From a theoretical point of view the ability to simulate the theory enables
tests of many conceptual issues related to confinement and chiral symmetry
breaking in the entire $\mu_I$-$T$ phase diagram, as we set out to show below.

Staggered fermions are often used for such investigations due to their remnant
chiral symmetry. Kogut-Sinclair \cite{ks} introduced the corresponding fermion
action to investigate also whether the isospin symmetry is spontaneously 
broken:

\begin{equation}
S_F = \sum_{sites} \bar \chi ~[ \slashed{D}(\tau_3 \mu_I) + m + i \lambda_I
\epsilon \tau_2 ]~\chi~.
\end{equation}

Here $\chi, \bar \chi$ are two component flavour spinors, $\tau_i$ are the
$SU(2)$ flavour generators, $\epsilon = (-1)^{x+y+z+t}$ is the `$\gamma_5$' for
staggered fermions, $\mu_I$ and $m$ are isospin chemical potential and quark
mass respectively and $\lambda_I$ is a pionic source that is sent to zero at the
end of the analysis. Ref. \cite{ks} worked out the symmetry-breaking patterns
and the corresponding observables which signal them. Further, it was argued that
the fermion determinant is positive definite, enabling simulations. 

\begin{figure}[htb]
\includegraphics[scale=0.20] {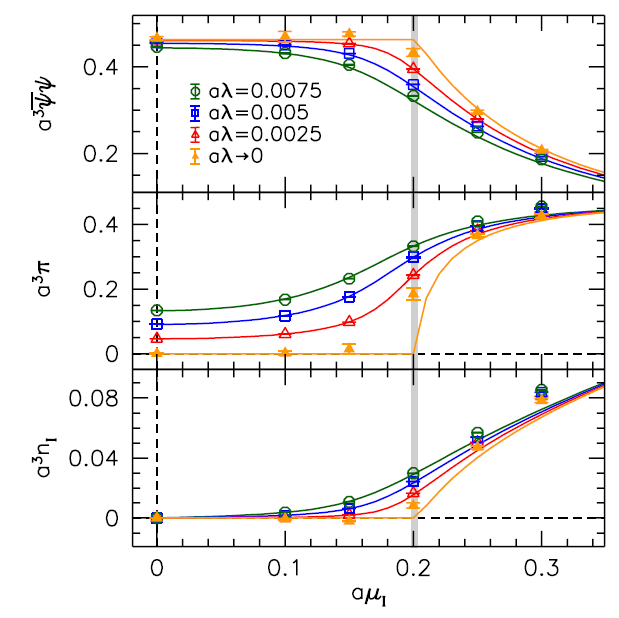}
\includegraphics[scale=0.25] {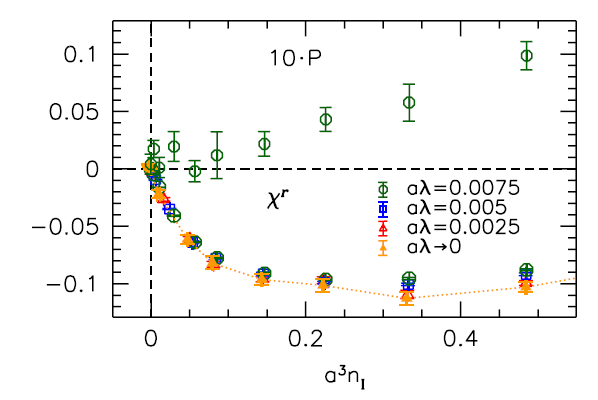}
\caption{Results for the chiral condensate, pion condensate, isospin density
(left panel) and Polyakov line (upper right panel) on $8^4$ lattice from 
Ref \cite{ge} for $m_\pi \simeq$ 260 MeV.}
\label{geth}
\end{figure}

Employing staggered fermions on $8^4$ lattices with $a =0.299(2)$ fm at a
lattice quark mass $ma =0.025$, corresponding to $m_\pi \simeq$ 260 MeV, 
Endr{\H o}di \cite{ge} recently investigated the phase structure. As can be seen
from his results in Figure \ref{geth} on the chiral condensate, the pion 
condensate, the isospin density and the Polyakov line, obtained from the
partition function $Z$ defined by $S_F$ above and the Wilson gluonic action 
by using, $ \langle \bar \psi \psi \rangle = \frac{T}{V} \frac {\partial~log Z}
{\partial m}$, $\langle \pi \rangle = \langle \bar \psi_u \gamma_5 \psi_d - 
\bar \psi_d \gamma_5 \psi_u \rangle = T {\partial~log Z}/{ V\partial \lambda_I}$,
$\langle n_I \rangle = T {\partial~log Z}/{V \partial \mu_I}$,
suggest a $a\mu^c_I \simeq 0.2$ in the $\lambda_I \to 0$ limit.

A linear $\lambda_I \to 0$ extrapolation of the data for the three $\lambda_I$
values indicated is displayed by points whereas the corresponding line is a
chiral theory fit. The grey vertical band denotes the value of $m_\pi/2$ in the
lattice units. The chiral condensate drops rapidly  around $\mu^c_I \simeq 
m_\pi/2$, where the pion condensate and isospin density also become nonzero
as does the Polyakov loop displayed in the upper half of the right panel. These
results shows deconfinement to occur as well with the chiral symmetry
restoration at the transition point, $\mu^c_I$.  The similarity of this
phenomenon with the finite temperature transition, i.e, $\mu_B = 0 = \mu_I$,
prompts further investigation of the nature of this transition in terms the
established ideas, such as topological excitations, or phenomenological models,
such the well-known Instanton-liquid model \cite{ssd} built on Instanton-fermion
couplings. 

Lattice QCD simulations support for the model was observed in the peak of the
Instanton-distribution at a radius $\rho =0.3$ fm \cite{DeHa}. Note that Overlap
Dirac operator, which has {\em exact} chiral symmetry on the lattice as well as
an index theorem, was used for this analysis, by studying its low eigenmodes
spectrum. Such studies were also carried out for the high temperature phase.
Number of low eigenmodes were found to get depleted as $T$ increased away from
$T_c$ \cite{ehkn,ggl}.  Furthermore a gap appeared to separate the low modes
from others.  Localized zero modes were observed \cite{ggl} for 1.25 $\le T/T_c
\le$ 2, suggesting the axial symmetry group $U_A(1)$  to be restored only
gradually up to 2$T_c$. Indeed, the scalar\& pseudoscalar meson correlators were
equal, as expected in a chiral symmetry restored phase, only after the
contribution of these zero modes was subtracted out from the former.  Clearly,
a similar investigation will be interesting for the nonzero chemical potential
case as well, in view of the both the na{\i}ve model expectations and the
results in Figure \ref{geth} for $\mu_I \ne 0$.

\section{Our Results}

Employing dynamical configurations on $24^3 \times 6$ lattices, generated with
a Symanzik improved action with 2  stout steps and for a quark mass tuned to 
have the physical pion mass, we investigated the the eigenvalue spectra of the
Overlap Dirac operator both below and above the isospin breaking phase
transition at $a \mu^c_I =0.1 $, which again corresponds to $\mu^c_I$ being 
$m_\pi/2$.  We employed the Arnoldi method to extract the eigenvalues of Overlap
Dirac operator, demanding a residue $ r = || DX -\eta || \le 10^{-10}$. It may
be noted that the dynamical configurations are with a nonzero $\mu_I= 0.05$ and
0.15, but there is no explicit $\mu_I$ in the operator itself, since our
intention is to study the topological fluctuations. We extracted $\sim$500 
eigenvalues from each configuration. At both the $\mu_I$ values computations
were done for two different values of $\lambda_I$ --- the isospin breaking 
parameter in the quark matrix.

\begin{figure}[htb]
\hskip 1.5 cm \includegraphics[scale=0.35] {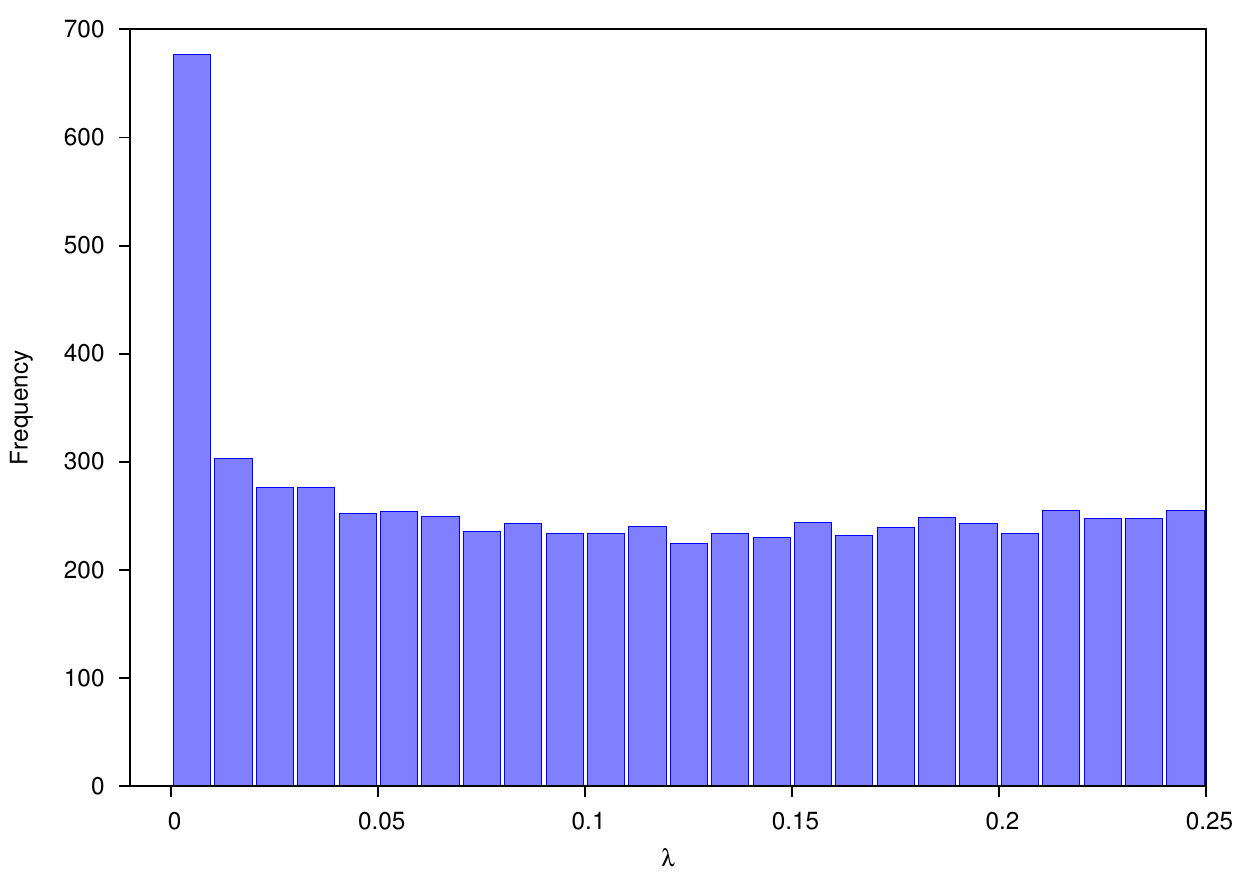}
\includegraphics[scale=0.35] {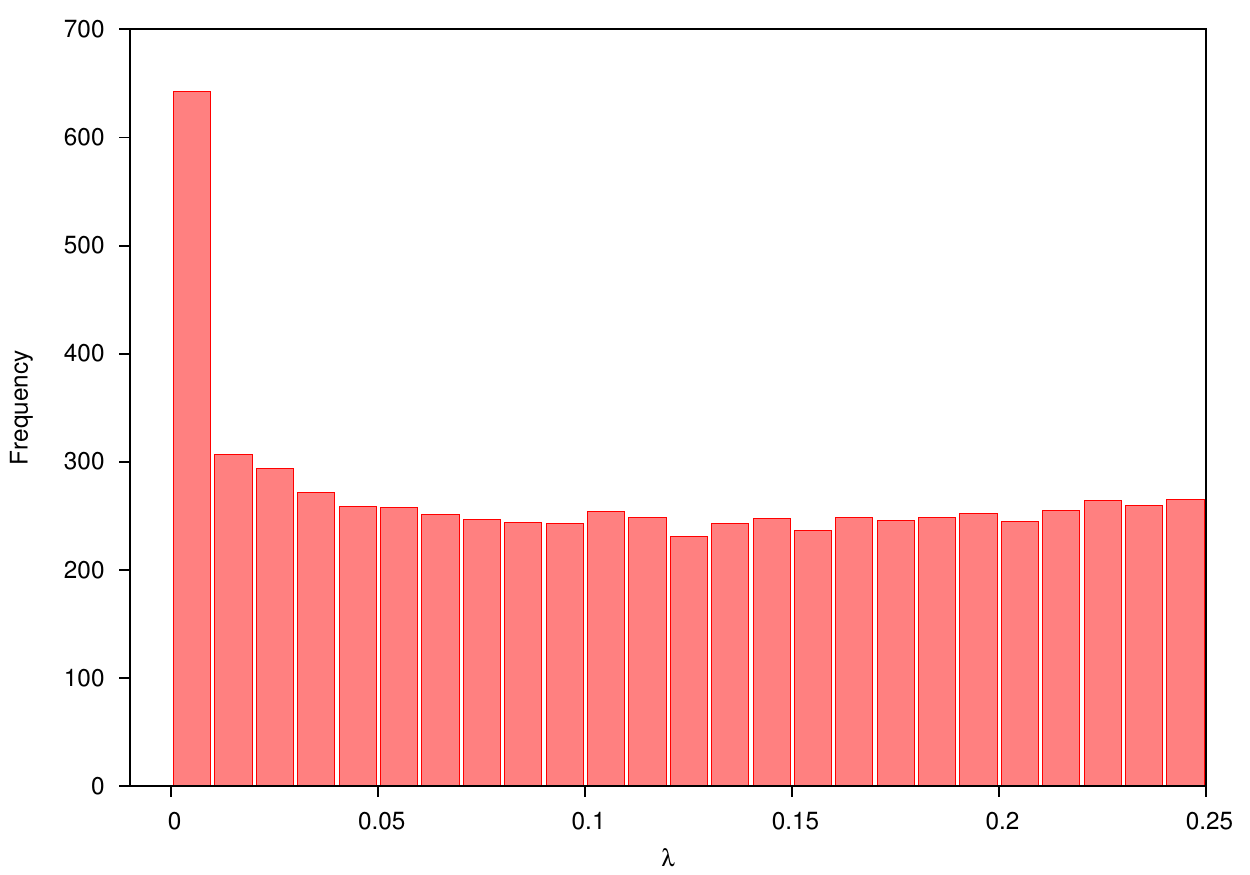}
\caption{Eigenvalue spectrum of the Overlap Dirac operator on $24^3 \times 6$
lattice for $\lambda_I=0.0006$ and $\mu_I/\mu^c_I =0.5$ (left panel) and 1.5 
(right panel).}
\label{lams}
\end{figure}

Since the eigenvalue $\lambda$ is complex for $D_{ov}$, we display in Figure
\ref{lams} the $|\lambda|$-distributions for $\lambda_I=0.0006$ both below and
above the transition. Fairly uniform distribution with some low modes are seen
in both the cases. Surprisingly the distributions are very similar as well, and
by overlaying them one finds them almost indistinguishable with minor
quantitative differences. Re-plotting them on a log scale, one can
easily identify the zero modes from the gap in the spectrum. Explicit chirality
checks were made to confirm their nature. Zooming in on the eigenvalue 
\begin{figure}[htb]
\hskip 1.5 cm \includegraphics[scale=0.35] {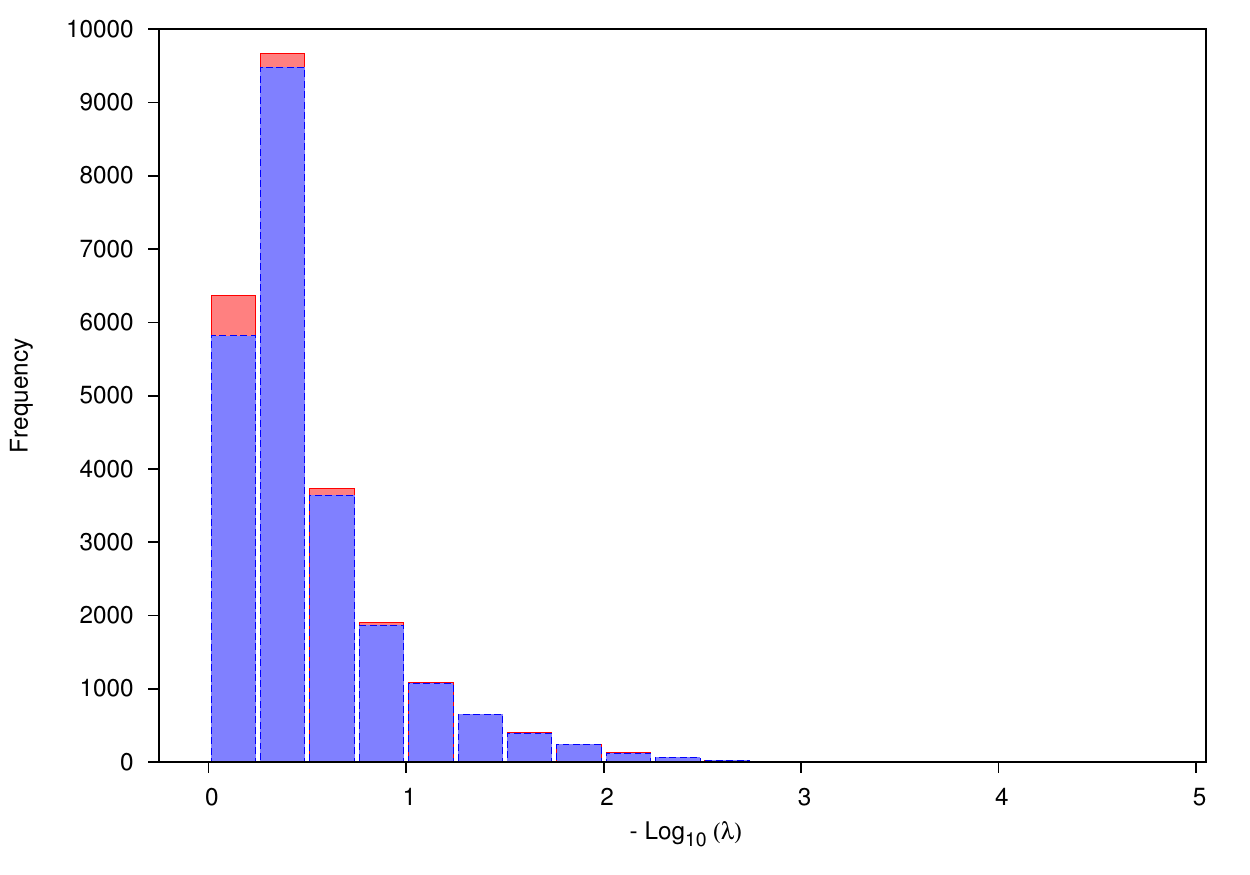}
\includegraphics[scale=0.35] {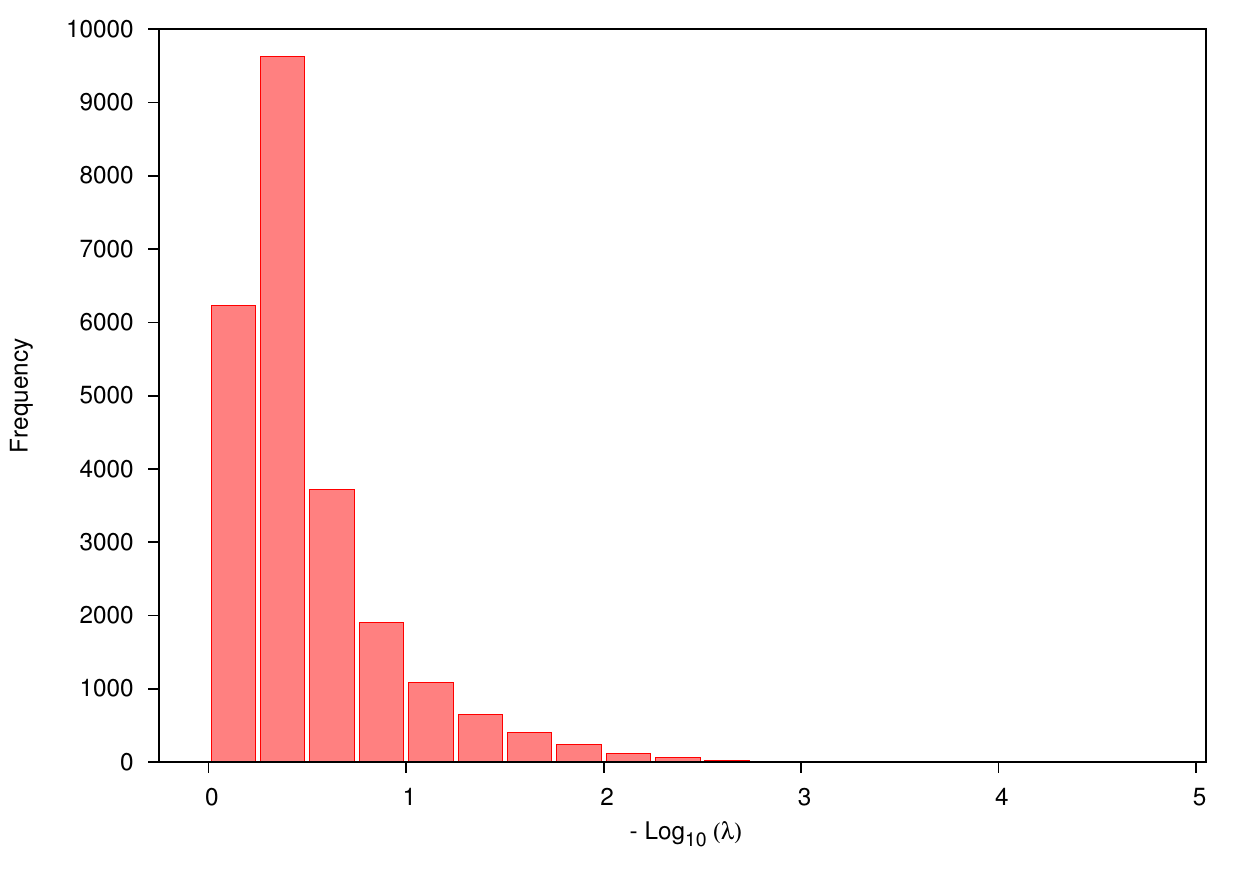}
\caption{Near-zero eigenvalue spectrum of the Overlap Dirac operator on 
$24^3 \times 6$ lattice for $\lambda_I=0.0025$ for both the $\mu_I$ values
(left panel) and for $\lambda_I=0.0006$ with $\mu_I/\mu^c_I =1.5$ 
(right panel).}
\label{lams1}
\end{figure} 
distribution on the log scale one can see if the near-zero modes have any
visible differences. While a nice smooth rise is seen in Figure \ref{lams1}, as
one moves away from the zero eigenvalue, the similarity in the distributions for
$\mu_I/\mu^c_I =0.5$ and 1.5 persists for even higher $\lambda_I$=0.0025, 
where one observes qualitatively the same picture as discussed above.
The displayed overlay of  near-zero modes for $\mu_I/\mu^c_I=0.5$ and 1.5 for
$\lambda_I=0.0025$ in the left panel also compares well with that of
$\mu_I/\mu^c_I=1.5$, $\lambda_I=0.0006$ in the right panel.

The exact chiral symmetry of the overlap fermions implies that nonzero modes are
doubly degenerate with opposite chirality while the zero modes possess only a
specific chirality. The latter act as a measure of topology due to the index
theorem the overlap fermions satisfy.  Table \ref{cmpr} list the number of zero
modes we observed in a sample of 50 independent gauge configurations as a
function of $\mu_I$ and $\lambda_I$. The last two columns list the corresponding
results of Ref.  \cite{ggl} which are also on samples of 50 configurations but
as a function of temperature in the vicinity of the finite temperature
transition at $\mu =0$.  While a steep fall of is seen in the latter as a
function of $T/T_c$, almost no variation is observed across $\mu_I$ for 
$\lambda_I =0.0006$ and a mild one for $\lambda_I =0.0025$, $\sim$25\% reduction.

\begin{table}[h]
\begin{center}
\caption{Number of zero modes $N^\lambda$ as a function of $\mu_I$ and
$\lambda_I$ along with corresponding results for finite temperature from Ref.
\cite{ggl}.}
\smallskip
\begin{tabular}{|c|c|c||c|c|}
\hline
$\mu_I/\mu^c_I$ & $N^{0.0006}_{zero}$ & $N^{0.0025}_{zero}$&$T/T_c$ &
$N_{zero}$ \\
\hline 
0.5  &  426 & 416 &1.25  &   18 \\
\hline
1.5  &  451 & 310 & 1.5  &    8 \\
\hline
-&-&-&2.0  &  1 \\
\hline
\end{tabular}  
\label{cmpr}
\end{center}
\end{table}  
\vskip -2.5 cm

\section{Summary}

We investigated the eigenvalue distribution for chirally exact Overlap Dirac
operator for $\mu_I/\mu^c_I =$ 0.5 \& 1.5, {\em i.~e.}, below and above the
isospin phase transition, which is indicated \cite{ge} to be similar to the
finite temperature transition in having both chiral symmetry restoration and a
rise of the Polyakov loop at the transition point. The distribution of zero and
near-zero modes is nearly the same for both at $\lambda_I$ =0.0006, with a 25 \%
reduction in former at $\lambda_I$ =0.0025.

This should be contrasted with the earlier $T\ne 0$ results \cite{ehkn, ggl},
where too these modes were present above the transition but decreased sharply as
one moved away from the transition. Further quantitative investigations in
pinning down the changes in these modes may help in efforts to understand the
difference in $T$ and $\mu_I$ directions, if any.

\section{Acknowledgements}
This work was supported by the Alexander von Humboldt foundation under its
Institutspartnerschaft Regensburg--Mumbai project. We gratefully acknowledge its
financial support. Two (RVG \& NM) of us thank ILGTI, TIFR, Mumbai for its
support and G.\ Endr\H{o}di acknowledges support from the DFG (Emmy Noether
Programme EN 1064/2-1).

\end{document}